\documentclass{emulateapj}

\shorttitle{A Bipolar Molecular Outflow from a Young Brown Dwarf}
\shortauthors{Phan-Bao et al.}

\begin{document}

\title{First Confirmed Detection of a Bipolar Molecular Outflow from a Young Brown Dwarf}

\author{Ngoc Phan-Bao,\altaffilmark{1}
Basmah Riaz,\altaffilmark{2}
Chin-Fei Lee,\altaffilmark{1}
Ya-Wen Tang,\altaffilmark{1}
Paul T.P. Ho,\altaffilmark{1,3}
Eduardo L. Mart\'{\i}n,\altaffilmark{2,4}
Jeremy Lim,\altaffilmark{1}
Nagayoshi Ohashi,\altaffilmark{1}
Hsien Shang\altaffilmark{1}}
\altaffiltext{1}{Institute of Astronomy and Astrophysics, Academia Sinica, P.O. Box 23-141, 
Taipei 106, Taiwan, ROC; pbngoc@asiaa.sinica.edu.tw}
\altaffiltext{2}{Instituto de Astrof\'{\i}sica de Canarias, 
C/ V\'{\i}a L\'actea s/n, E-38200 La Laguna (Tenerife), Spain.}
\altaffiltext{3}{Harvard-Smithsonian Center for Astrophysics, Cambridge, MA}
\altaffiltext{4}{University of Central Florida, Dept. of Physics, 
PO Box 162385, Orlando, FL 32816-2385}

\begin{abstract}
Studying the earliest stages in the birth of stars is crucial
for understanding how they form.
Brown dwarfs with masses between that of stars and planets
are not massive
enough to maintain stable hydrogen-burning fusion reactions during most
of their lifetime.
Their origins are subject to much debate in recent literature because their
masses are far below the typical mass where core collapse is expected to occur.
We present the first confirmed evidence that brown dwarfs undergo a phase of molecular
outflow that is typical of young stars. Using the Submillimeter Array, 
we have obtained a map of a bipolar molecular outflow from
a young brown dwarf. 
We estimate an outflow mass of $1.6 \times 10^{-4} M_{\odot}$ and a mass-loss rate of
$1.4 \times 10^{-9} M_{\odot}$. These values are over two orders of magnitude
smaller than the typical ones for T Tauri stars.
From our millimiter continuum data and our own analysis of Spitzer infrared 
photometry, we estimate that the brown dwarf has a disk with a mass 
of $8 \times 10^{-3} M_{\odot}$ and an outer disk radius of 80 AU. 
Our results demonstrate that the bipolar 
molecular outflow operates down to planetary masses, occurring in
brown dwarfs as a scaled-down version of the universal process seen in young stars.
\end{abstract}


\section{Introduction}
Star formation starts with collapse, accretion and launching of material
as a bipolar outflow \citep{lada}.
It is thought that brown dwarfs could 
undergo the same stages of formation as stars and that they have a common 
origin \citep{luhman07}. 
However the typical masses of brown dwarfs, 15--75~$M_{\rm J}$, 
are far below 
the typical Jeans mass in molecular clouds, and hence it is difficult to make a large 
number of them by direct gravitational collapse. Several brown dwarf formation 
mechanisms have been proposed (see \citealt{whitworth}
and references therein), and more observations are needed to 
improve our understanding of the role of each of them.

Recent observational evidences for accretion, and optical jets 
have been found in
young brown dwarfs \citep{martin01,fer01,jay,whelan07}.
However there is no firm evidence up to now to demonstrate that the brown dwarf formation
involves the bipolar molecular outflow process
as seen in young stars.
In the case of low-mass stars, 
the molecular outflow plays a vital role 
\citep{ba92,ba96,rei01,mckee}
to testing the low-mass star formation theory \citep{shu91,masson,raga,pudritz}. 
Therefore, the detection of molecular outflows provides a key piece of 
information on the earliest phase of the brown dwarf formation and allows us
to constrain the low-mass star formation theory in the sub-stellar domain.

In this Letter, we present millimeter observations 
of the young brown dwarf ISO-Oph~102 in the $\rho$ Ophiuchi dark cloud. 
We detect a 
bipolar molecular outflow from the brown dwarf, demonstrating that this outflow mechanism 
as seen in young stars continues to operate down to the masses of brown dwarfs.
We show that the outflow mass is much smaller than typical values found 
in low-mass stars by two to three orders of magnitude. 
We also combine our radio data with infrared data to estimate disk parameters.
The infrared spectrum of ISO-Oph~102 
shows strong crystalline silicate features. This demonstrates
dust processing in the protoplanetary disk \citep{apai} of the brown dwarf.
The coexistence of the molecular outflow with the crystallization of dust grains 
could be of importance for models of planet formation. 

Sec.~2 presents the millimeter observations and the data reduction,
Sec.~3. discusses the physical properties of the bipolar molecular outflow,
the disk parameters, and the crystalline silicate features in 
ISO-Oph~102, Sec.~4 summarizes our results.
\section{Observations and data reduction}
ISO-Oph~102 (or [GY92]~204), a young M6 dwarf
initially identified by \citet{greene}, is 
located in the $\rho$ Ophiuchi dark cloud at a distance of 
125~parsecs \citep{degeus89}.  
Its estimated mass is 60~$M_{\rm J}$ \citep{natta02}.
This value is below the Hydrogen-burning limit and in good agreement
with the dynamically measured mass one of an M6.5 dwarf of 55~$M_{\rm J}$ at the same age
(see \citealt{stassun}). ISO-Oph~102 is therefore a young
bona-fide brown dwarf.
The presence of the Lithium absorption in its optical spectrum \citep{natta04}
supports its brown dwarf nature \citep{martin94}. 
Strong evidences for the presence of an accretion disk
have been reported \citep{natta04}. The blue-shifted optical jet
component was discovered by \citet{whelan05}.
The optical visibility of $\rho$~Oph~102 suggests that the source 
corresponds to a class II object \citep{lada} in the star formation phase, 
a class with an accreting circumstellar
disk and the protostar at this stage is the so-called 
classical T Tauri star.

We have observed ISO-Oph~102 with the receiver band at 230~GHz of the SMA\footnote{
The Submillimeter Array is a joint project between the 
Smithsonian Astrophysical Observatory and 
the Academia Sinica Institute of Astronomy and Astrophysics 
and is funded by the Smithsonian Institution and the Academia Sinica.} \citep{ho}. 
Both 2 GHz-wide sidebands which are separated
by 10 GHz were used. The SMA correlator was configured with high spectral
resolution bands of 512 channels per chunk of 104~MHz for $^{12}$CO, $^{13}$CO, 
and C$^{18}$O $J=2 \rightarrow 1$ lines, giving a channel spacing of 0.27~km~s$^{-1}$.
A lower resolution of 3.25~MHz per channel was set up for the remainder of
each sideband. The quasars 1625$-$254 and 3C~279 have been observed for gain
and passband calibration, respectively. Mars were used for
flux calibration. The data were calibrated using the MIR software package
and further analysis was carried out with the MIRIAD package adapted
for the SMA. 
The compact configuration was used, resulting a synthesized beam of 
3$''$.60~$\times$~2$''$.43 with a position angle of 8.5$^{\circ}$. 
The rms (root mean square) sensitivity was about 3~mJy for the continuum,
using both sidebands and $\sim$0.2~Jy~beam$^{-1}$ per channel for
the line data. The primary FWHM (full width at half maximum) 
beam is about 55$''$~at the observed frequencies.
The integrated flux density from the dust continuum emission was 7$\pm$3~mJy
measured at the brown dwarf position.
\section{Discussion}
An overlay of a near-infrared image and 
the integrated intensity in the carbon monoxide (CO $J=2-1$) line 
emission is shown in Figure 1.
Two spatially resolved blue- and red-shifted
CO components are symmetrically displaced
on opposite sides of the brown dwarf position, with
the size of each lobe of about 8$''$ corresponding
to 1000~AU in length.
This is similar to the typical pattern of bipolar molecular outflows
as seen in young stars \citep{lada}. 
The two outflow components with a wide range of velocity (Figure 2)
suggests a bow shock structure, an effect of the interaction
between the jet propagation and the ambient material, which appears very similar
to the bow shock phenomena as seen in young stars \citep{lee}.
Such a CO outflow morphology suggests that
the jet-driven bow shock model (e.g., \citealt{masson}) may be at work in $\rho$~Oph~102.

Following the standard manner \citep{cabrit,andre}, we calculate the
outflow properties. We use a value of 35~K \citep{loren}
for the excitation temperature, and we derive a lower limit to 
the outflow mass of $3.2 \times 10^{-5} M_{\odot}$. If we correct for optical depth
with a typical value of 5 \citep{l88a}, we obtain an upper limit
to the outflow mass of $1.6 \times 10^{-4} M_{\odot}$. This
value is smaller than the ones observed in young stars \citep{l88b}
by three orders of magnitude.

To estimate the outflow inclination, we combine our SMA data with archival 
infrared data of the Spitzer Space Telescope. 
For the infrared photometry, we use the archival basic calibrated data (PIDs: 58, 177)
that were reduced with IRAF\footnote{IRAF is distributed by the 
National Optical Astronomy Observatories,
which are operated by the Association of Universities for Research
in Astronomy, Inc., under cooperative agreement with the National
Science Foundation.} 
to measure flux densities.
Disk modeling is performed 
as the previous works (see \citealt{riaz} and references therein). 
The circumstellar geometry consists of a rotationally flattened infalling envelope, 
bipolar cavities, 
and a flared accretion disk in hydrostatic equilibrium. 
The disk density is proportional to $\varpi^{-\alpha}$, 
where $\varpi$ is the radial coordinate in the disk midplane, 
and $\alpha$ is the radial density exponent. 
The disk scale height increases with radius, $h=h_{0}(\varpi / R_{*})^{\beta}$, 
where $h_{0}$ is the scale height at $R_{*}$ (the brown dwarf radius)  
and $\beta$ is the flaring power. 
Stellar parameters are obtained from \citet{natta04}, 
and an accretion rate of $10^{-9} M_{\odot}$~yr$^{-1}$ is used. 
The NextGen \citep{haus} atmosphere file for an effective temperature 
$T_{eff}$ of 2700~K, 
and log {\it g} = 3.5, where $g$ is the surface gravity,
is used to fit the atmosphere spectrum of the central sub-stellar source. 
The disk shows some flaring longward of $\sim$10 $\mu$m, 
and the best model-fit was obtained using a disk mass of $8 \times 10^{-3} M_{\odot}$, 
an inclination angle between 63$^{\circ}$ and 66$^{\circ}$, 
an outer disk radius of 80 AU, and an inner disk radius equal to $R_{\rm sub}$, where $R_{\rm sub}$
is the dust sublimation radius: 1~$R_{\rm sub}$$\sim 6.8~R_{\odot}$.
There is thus no inner hole in the disk as it would have to be larger than $R_{\rm sub}$. 
The outer disk radius and the disk mass are well-constrained by the millimeter observations. 
A more face-on inclination can be ruled out as it results in larger-than-observed mid- and 
far-infrared fluxes. 
Figure 3 presents the best fit
from the disk modeling obtained using a disk mass of $8 \times 10^{-3} M_{\odot}$, 
an inclination angle between 63$^{\circ}$ and 66$^{\circ}$, 
an outer disk radius of 80 AU, resulting in a projected disk radius
of 32-36 AU. 
This immediately explains the non-detection of the red-shifted optical jet component, 
estimated to be about 15~AU in length as reported
in the previous observation \citep{whelan05}. Since the projected disk radius is larger
than the jet length, the disk therefore hides the red-shifted jet component
from our view.

We use the observed maximum flow velocity of 2.2 km~s$^{-1}$ and apply
a correction for the outflow inclination to compute upper limit
values for the kinematic and 
dynamic parameters. 
We find that the momentum is $P=1.4 \times 10^{-4}$$M_{\odot}$~km~s$^{-1}$,
the energy is $E=3.1 \times 10^{-4}$$M_{\odot}$~km$^{2}$~s$^{-2}$, 
the force is $F=5.9 \times 10^{-8}$$M_{\odot}$~km~s$^{-1}$~yr$^{-1}$,
and the mechanical luminosity is $L=2.1 \times 10^{-5}$~L$_{\odot}$, where L$_{\odot}$
is the solar luminosity. A correction for the optical depth factor of 5 and 
a missing flux factor of 3 for SMA \citep{bourke} will increase these upper limit values 
by a factor of 15. 
Using the lower value of the outflow mass and
a correction factor of 10 applied for the outflow duration time \citep{parker},
we derive the outflow mass-loss rate of 
$1.4 \times 10^{-9}$$M_{\odot}$~yr$^{-1}$, which is smaller
than a typical value for T Tauri stars \citep{lada} by two orders of magnitude.
One should note that
these values are roughly estimated, depending on the correction factor values
used in the calculation.
One should also note that the outflow mass and the mass loss rate values from ISO-Oph 102 appear
similar to that from L1014-IRS \citep{bourke}, suggesting that they share
the same origin. However as clearly stated in \citet{huard} L1014-IRS could be an embedded
protostar or a proto-brown dwarf, corresponding to the earliest stages of star 
or brown dwarf formation processes, respectively. 
More observations are needed to confirm the L1014-IRS nature and 
study the connection between ISO-Oph 102 and L1014-IRS
in the whole formation phase.

We also examine the possibility that the emission might be due to bound motions
and not outflow emission. This would require
an interior mass \citep{lada} of 2.7$M_{\odot}$ for an outflow size of
1000~AU with a velocity of 2.2~km~s$^{-1}$, which is significant larger
than the core mass of $< 0.4$~$M_{\odot}$ within the same radius \citep{motte,young}.
We therefore conclude that the detected emission is from the outflow.

The IRS infrared (7.5-14.3 $\mu$m) \citep{houck}
spectra of ISO-Oph~102 
whose the basic-calibrated data were downloaded from the Spitzer archive (PID: 3499)
and reduced using the SMART software package \citep{higdon}
are also analyzed. It is worthy to note here that 
the mid-infrared spectrum of ISO-Oph~102  shows crystalline
silicate features: enstatite (MgSiO$_{3}$) at 9.3~$\mu$m and very strong 
forsterite (Mg$_{2}$SiO$_{4}$) at 11.3~$\mu$m (see Figure 4). The brown dwarf is indeed
an analog of SST-Lup3-1, which is an M5.5 brown dwarf in the Lupus III dark cloud \citep{merin}. 
The crystalline contribution to the silicate feature (flux at 11.3 $\mu$m
over flux at 9.8 $\mu$m) in ISO-Oph~102 is 1.1, even greater than 
the value of 0.9 in SST-Lup3-1 \citep{merin}.
This provides a direct evidence of grain growth and dust settling,
indicating the object is in the transition phase between
the class II and III (a class with an optically thin disk) and the brown
dwarf is reaching the final mass. 

\section{Summary}
Here, we report the first confirmed detection of the bipolar molecular outflow
in young brown dwarfs. We characterize the fundamental properties
of the outflow with the first insight on its morphology.
The results indicate that the bipolar molecular outflow in brown dwarfs is very similar
to outflows as seen in young stars but scaled down by three and two 
orders of magnitude for the outflow mass and the mass-loss rate, respectively.
We demonstrate for the first time that brown dwarfs,
even young planetary mass
objects \citep{martin01} can launch a  
bipolar molecular outflow, and hence we confirm that they are likely to have 
a common origin with the low-mass stars. 
This suggests that the terminal stellar/brown dwarf mass 
is not due to different formation mechanisms, 
but more likely due to the initial mass of the cloud core. 
We also show that there is evidence 
for dust crystallization in the disk around the brown dwarf. 
In brown dwarfs the molecular outflows may be longer lived than in low-mass 
stars and they coexist with grain growth and crystallization. 
Outflows from brown dwarfs may play a crucial role in sweeping out gas from 
the disk and favouring the formation of rocky planets.

\acknowledgments
NP-B has been aided in this work by a Henri Chr\'etien International Research Grant administered 
by the American Astronomical Society.
This work is based in part on observations made with the Spitzer 
Space Telescope, which is operated by the Jet Propulsion Laboratory,
California Institute of Technology, under a contract with NASA.
This work has made use of the Centre de Donn\'ees astronomiques de Strasbourg (CDS) 
database.

\clearpage

\begin{figure}
\vskip 1in
\hskip -0.25in
\centerline{\includegraphics[width=6in,angle=0]{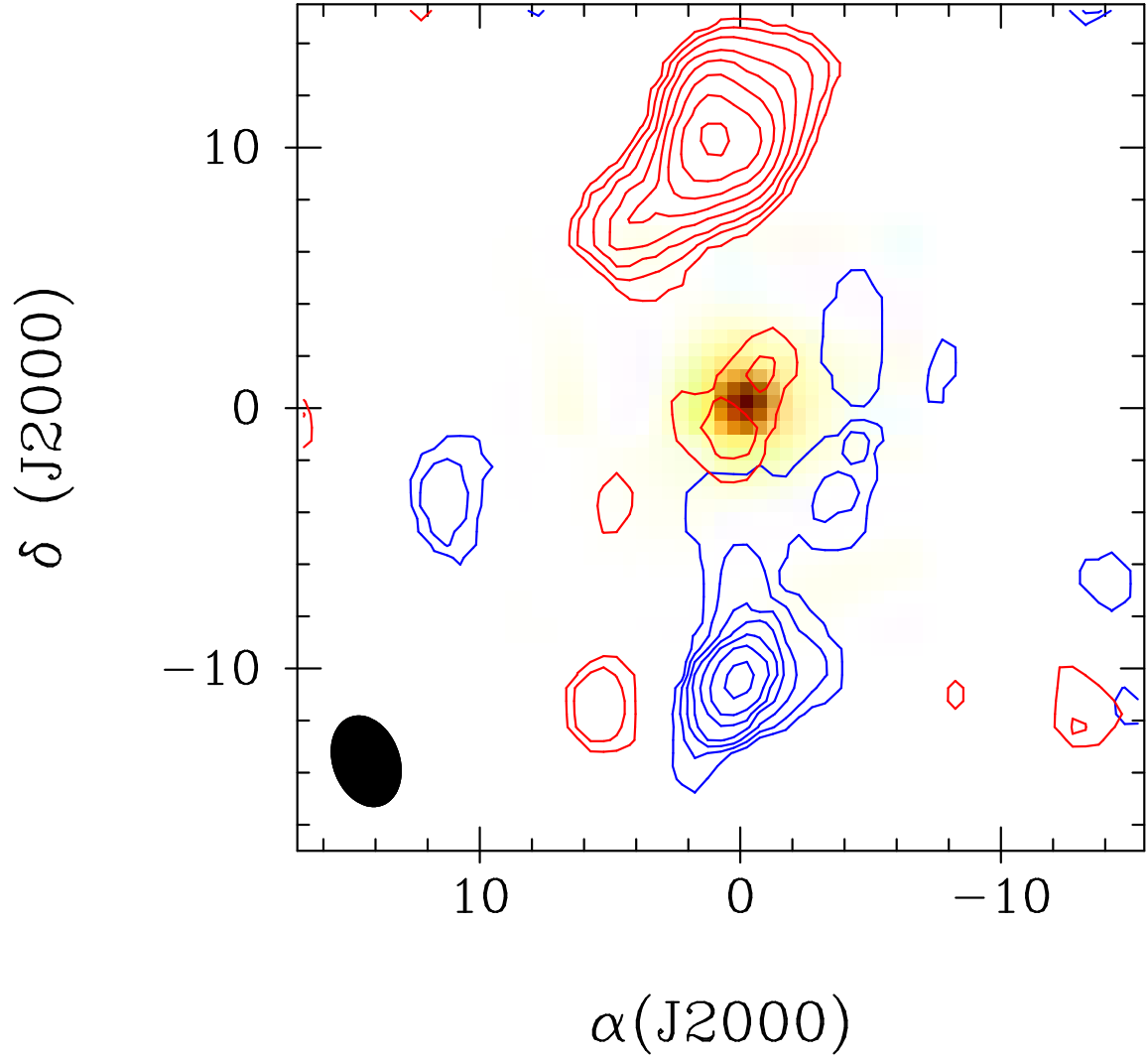}}
\caption{\normalsize An overlay of the J-band (1.25 $\mu$m) near-infrared 
Two Micron All Sky Survey (2MASS) image and
the integrated intensity in the carbon monoxide (CO $J=2-1$) line 
emission from 3.8 to 7.7~km~s$^{-1}$ line-of-sight velocities.
The blue and red contours represent the blue-shifted 
(integrated over 3.8 and 5.9~km~s$^{-1}$) and red-shifted (integrated over 5.9 and 7.7~km~s$^{-1}$)
emissions, respectively. 
The contours are 3,6,9,...times the 
rms of 0.15~Jy~beam$^{-1}$~km~s$^{-1}$. The brown dwarf is visible in the
J-band image. The position angle of the outflow is about 3$^{\circ}$. 
The peaks of the blue- and red-shifted components are symmetric to the center
of the brown dwarf with an offset of 10$''$. The synthesized beam is shown in the bottom
left corner.
\label{of}}
\end{figure}

\clearpage

\begin{figure}
\vskip 1in
\hskip -0.25in
\centerline{\includegraphics[width=5in,angle=-90]{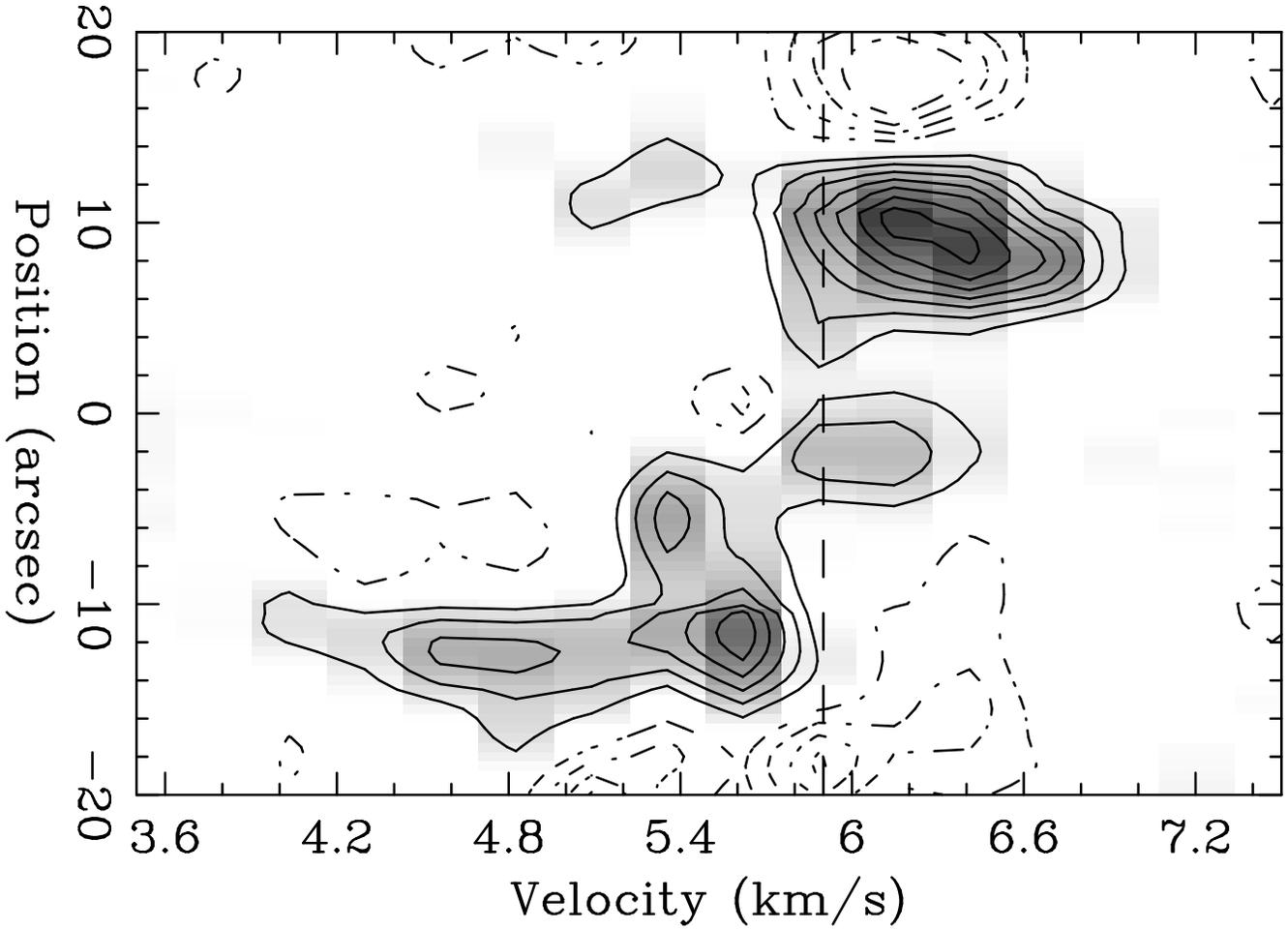}}
\caption{\normalsize Position-Velocity (PV) cut diagram for CO~$J=2 \rightarrow 1$ emission
at a position angle of 3$^{\circ}$.
The contours are $-$12, $-$9, $-$6, $-$3, 3, 6, 9, 12,...times the 
rms of 0.2 Jy beam$^{-1}$. The systemic velocity of the brown dwarf,
which is estimated by an average of the velocities 
of red- and blue-shifted components, is indicated
by the dashed line. Our value of 5.9$\pm$0.27~km~s$^{-1}$ is consistent with
the previously measured value \citep{whelan05} of 7$\pm$8~km~s$^{-1}$
within the error bar.
Both blue- and red-shifted components shows a wide range of
the velocity in their structure, which appears 
to be the bow-shock surfaces as observed in young stars \citep{lee}. These
surfaces are formed at the head of the jet and accelerate the material
in the bow-shock sideways (e.g., \citealt{masson}).
\label{pv}}
\end{figure}

\clearpage

\begin{figure}
\vskip 1in
\hskip -0.25in
\centerline{\includegraphics[width=6in,angle=0]{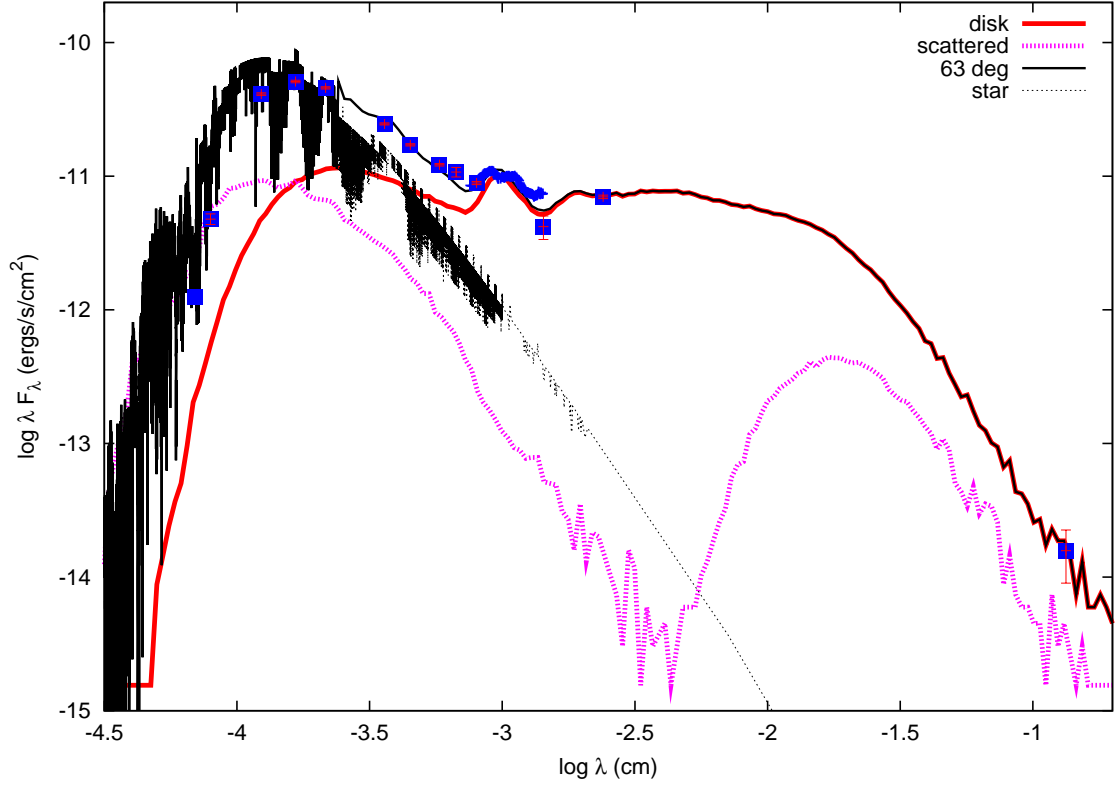}}
\caption{\normalsize Best fit model for ISO-Oph~102. Contributions
from the stellar photosphere, the disk and the scattered flux are indicated.
Optical and near-infrared data collected from the Vizier archive, infrared fluxes
from \citet{natta02} and
measured based on data from
the Spitzer archive (IRAC, MIPS, IRS),
millimeter data from our observations.
\label{sed}}
\end{figure}

\clearpage

\begin{figure}
\vskip 1in
\hskip -0.25in
\centerline{\includegraphics[width=5in,angle=-90]{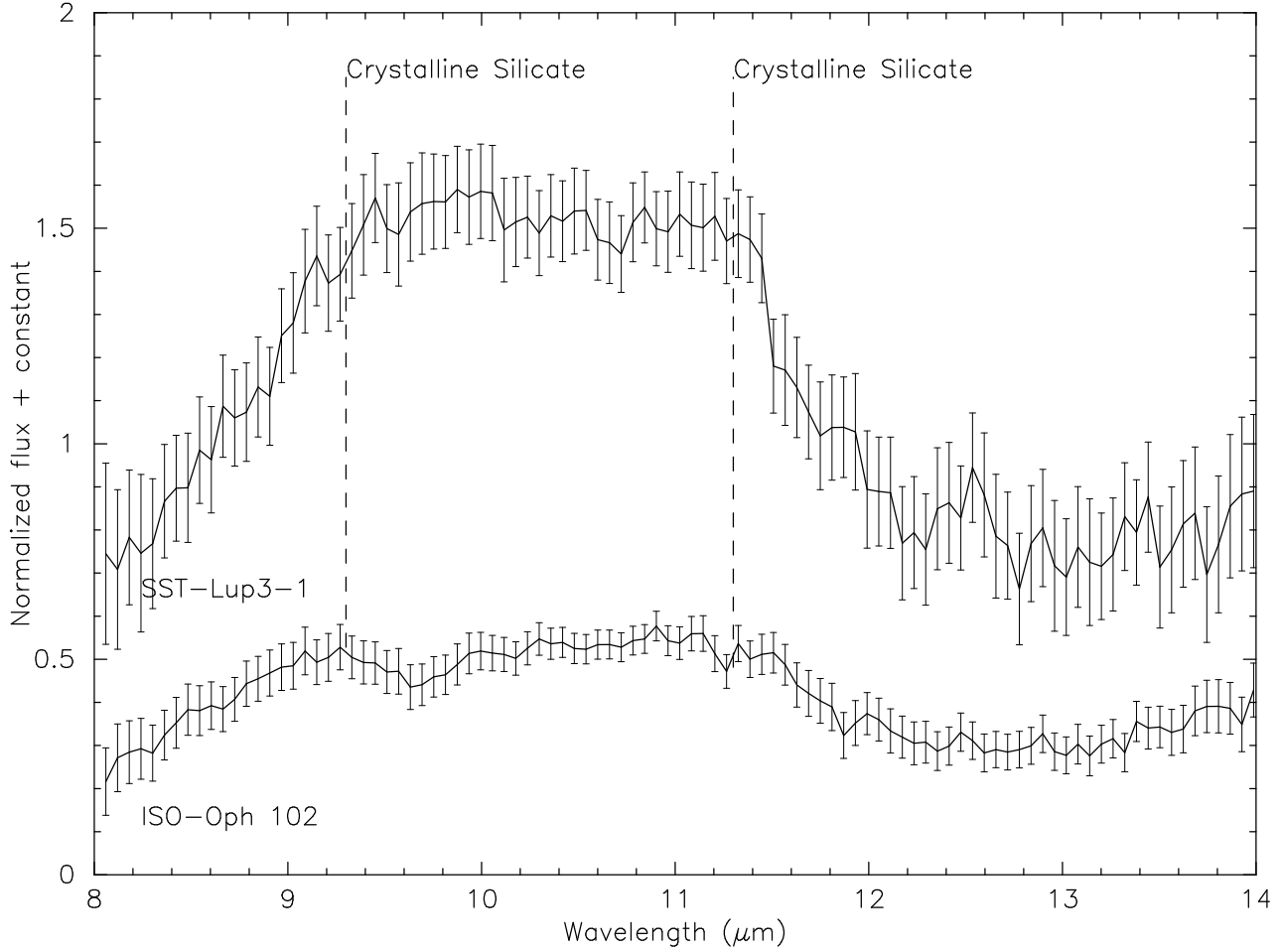}}
\caption{\normalsize Infrared spectral comparison between ISO-Oph 102 and
SST-Lup3-1 \citep{merin} whose spectrum obtained from the Spitzer archival data (PID: 179).
Continuum subtraction was done following the literature \citep{van03,apai}.
The crystalline silicate features at 9.3 $\mu$m (mainly enstatite) and 11.3 $\mu$m
(forsterite) are indicated. 
\label{cs}}
\end{figure}

\end{document}